\documentclass[12pt,letterpaper]{article}
\pdfoutput=1
\usepackage{jheppub}
\usepackage{amsfonts, amsthm}
\usepackage[english]{babel}
\usepackage[utf8]{inputenc}
\usepackage{slashed}
\usepackage{mathrsfs}
\usepackage{amssymb}
\usepackage{color}
\hypersetup{unicode}

\newcommand{\eq}{\begin{equation}}
\newcommand{\feq}{\end{equation}}
\newcommand{\eqn}{\begin{eqnarray}}
\newcommand{\feqn}{\end{eqnarray}}

\newcommand{\ma}[1]{\mbox{$\mathcal{#1}$}}

\newcommand{\masf}[1]{\mbox{$\mathsf{#1}$}}
\newcommand{\D}{{\rm d}}

\title{Symplectically invariant flow equations for $N = 2$, $D = 4$ gauged supergravity with hypermultiplets}

\author{Dietmar Klemm,}
\author{Nicol\`o Petri}
\author{and Marco Rabbiosi}

\affiliation{Dipartimento di Fisica, Universit\`a di Milano, and \\
INFN, Sezione di Milano, \\
Via Celoria 16, I-20133 Milano, Italy.}

\emailAdd{dietmar.klemm@mi.infn.it}
\emailAdd{nicolo.petri@mi.infn.it}
\emailAdd{marco.rabbiosi@mi.infn.it}
\preprint{IFUM-1046-FT}

\abstract{We consider $N=2$ supergravity in four dimensions, coupled to an arbitrary
number of vector- and hypermultiplets, where abelian isometries of the quaternionic hyperscalar
target manifold are gauged. Using a static and spherically or hyperbolically symmetric ansatz for the
fields, a one-dimensional effective action is derived whose variation yields all the equations of motion.
By imposing a sort of Dirac charge quantization condition, one can express the complete scalar potential
in terms of a superpotential and write the action as a sum of squares. This leads to first-order
flow equations, that imply the second-order equations of motion. The first-order flow turns out to
be driven by Hamilton's characteristic function in the Hamilton-Jacobi formalism, and contains
among other contributions the superpotential of the scalars. We then include also magnetic gaugings
and generalize the flow equations to a symplectically covariant form. Moreover, by rotating the charges
in an appropriate way, an alternative set of non-BPS first-order equations is obtained that corresponds
to a different squaring of the action. Finally, we use our results to derive the attractor
equations for near-horizon geometries of extremal black holes.}


\begin{document}
\maketitle
\flushbottom

\section{Introduction}

Black holes in gauged supergravity theories provide an important testground
to address fundamental questions of gravity, both at the classical and quantum level. Among these are
for instance the problems of black hole microstates, the final state of black hole evolution,
uniqueness- or no hair theorems, to mention only a few of them. In gauged supergravity, the solutions
often have AdS asymptotics, and one can then try to study these issues guided by the AdS/CFT
correspondence. A nice example for this is the recent microscopic entropy
calculation \cite{Benini:2015eyy} for the black hole solutions to $N=2$, $D=4$ Fayet-Iliopoulos
gauged supergravity constructed in \cite{Cacciatori:2009iz}. These preserve two real supercharges,
and are dual to a topologically twisted ABJM theory, whose partition function can be computed
exactly using supersymmetric localization techniques. This partition function can also be interpreted as
the Witten index of the superconformal quantum mechanics resulting from dimensionally reducing the
ABJM theory on a two-sphere. To the best of our knowledge, the results of \cite{Benini:2015eyy} represent
the first exact black hole microstate counting that uses AdS/CFT and that does
not involve an $\text{AdS}_3$ factor\footnote{Or geometries related to $\text{AdS}_3$, like those
appearing in the Kerr/CFT correspondence \cite{Guica:2008mu}.} with a corresponding two-dimensional
CFT, whose asymptotic level density is evaluated with the Cardy formula.

On the other hand, black hole solutions to gauged supergravity are also relevant for a number of recent 
developments in high energy- and especially in condensed matter physics, since they provide the
dual description of certain condensed matter systems at finite temperature, cf.~\cite{Hartnoll:2009sz}
for a review. In particular, models that contain Einstein gravity coupled to $\text{U}(1)$ gauge fields\footnote{The necessity of a bulk $\text{U}(1)$ gauge field arises, because a basic ingredient of
realistic condensed matter systems is the presence of a finite density of charge carriers.}
and neutral scalars have been instrumental to study transitions from Fermi-liquid to
non-Fermi-liquid behaviour, cf.~\cite{Charmousis:2010zz,Iizuka:2011hg} and references therein.
In AdS/condensed matter applications one is often interested in including a charged scalar
operator in the dynamics, e.g.~in the holographic modeling of strongly coupled
superconductors \cite{Hartnoll:2008vx}. This is dual to a charged scalar field in the bulk, that
typically appears in supergravity coupled to gauged hypermultiplets. These theories are thus
particularly appealing in an AdS/cond-mat context, and it would be nice to
dispose of analytic black hole solutions to gauged supergravity with hyperscalars turned on.

Up to now, the only known such solution in four dimensions was constructed recently
in \cite{Chimento:2015rra}\footnote{Numerical black hole solutions in four-dimensional gauged
supergravity with hypers were obtained in \cite{Halmagyi:2013sla}. Solutions that have ghost modes
(i.e., with at least one negative eigenvalue of the special K\"ahler metric) were found
in \cite{Hristov:2010eu}. In five dimensions, a singular solution of supergravity with gauging of the
axionic shift symmetry of the universal hypermultiplet was derived in \cite{Gutperle:2001vw}.},
by using the results of \cite{Meessen:2012sr}, where all supersymmetric
backgrounds of $N=2$, $D=4$ gauged supergravity coupled to both vector- and hypermultiplets were
classified. Such BPS solutions typically satisfy first-order equations that arise from vanishing fermion
variations, and that are much easier to solve than the full second-order equations of motion. 

In our paper we shall derive such a set of first-order equations for static and spherically (or hyperbolically)
symmetric black holes, that will however be more general
than that of \cite{Meessen:2012sr}, in two respects. First of all, we consider also magnetic gaugings
in order to restore symplectic covariance. Second, our equations are not necessarily tied to supersymmetry,
but arise from writing the action as a sum of squares, making essential use of the Hamilton-Jacobi
formalism. This allows us to extend our results beyond the BPS case, and has the advantage
to potentially describe also nonextremal black holes, by appropriately modifying the Hamilton-Jacobi
function that we use here.

While we were not yet able to provide such an extension to the nonextremal case, our first-order
system may still have applications in holographic modeling of condensed matter phenomena,
for instance to study quantum phase transitions like those appearing in the high-$T_c$ cuprates when
one dopes the $\text{CuO}_2$-layers with charge carriers at zero temperature.

The remainder of this paper is organized as follows: In the next section, we
briefly review $N=2$, $D=4$ gauged supergravity coupled to vector- and
hypermultiplets. In section \ref{sec:flow-eqns} we consider gaugings of abelian isometries of the
quaternionic hyperscalar target manifold, impose staticity and spherical or hyperbolic
symmetry on the fields, and derive a one-dimensional effective action from which all the equations of
motion follow. It is then shown that under some rather mild additional assumptions one can explicitely
solve the Hamilton-Jacobi equation. This leads to first-order flow equations that we subsequently
generalize to include also magnetic gaugings and to the non-BPS case. Our results represent an extension
of the recent work \cite{Cardoso:2015wcf}, where only flat horizons and purely electric gaugings were
considered. In section \ref{sec:attractors} we plug the near-horizon geometry
$\text{AdS}_2\times\Sigma$ (where $\Sigma$ is a two-dimensional space of constant curvature)
into our system of first-order equations, and derive the symplectically covariant attractor equations
for gauged supergravity with hypermultiplets. Section \ref{sec:examples} contains some examples
of explicit solutions to the flow equations with running hyperscalars for models with the universal
hypermultiplet and one vector multiplet. We conclude in \ref{sec:final} with some final remarks.

\section{\label{Setup}Matter-coupled $N=2$, $D=4$ gauged supergravity}

The supergravity multiplet of $N=2$, $D=4$ supergravity can be coupled to a number $n_V$ of vector multiplets and to $n_H$ hypermultiplets. The 
bosonic sector then includes the vierbein $e^a{}_\mu$, $n_V+1$ vector fields $A^\Lambda_\mu$ with $\Lambda=0,\dots n_V$ (the graviphoton plus $n_V$ other fields from the vector multiplets), $n_V$ complex scalar fields $z^i$ ($i=1,\dots,n_V$), and $4 n_H$ real 
hyperscalars $q^u$ ($u=1,\dots,4n_H$).

The complex scalars $z^i$ of the vector multiplets parametrize an $n_V$-dimensional special K\"ahler manifold, i.e., a K\"ahler-Hodge 
manifold, with K\"ahler metric $g_{i\bar\jmath}(z,\bar z)$, which is the base of a symplectic bundle with the covariantly holomorphic 
sections\footnote{We use the conventions of \cite{Andrianopoli:1996cm}.}
\begin{equation}
 \ma{V}=\left(\begin{array}{c}
                   L^\Lambda\\
                   M_\Lambda
                  \end{array}\right), \qquad 
                  D_{\bar \imath}\ma{V}\equiv\partial_{\bar \imath}\ma{V}
                  -\frac{1}{2}\left(\partial_{\bar \imath}\ma{K}\right)\ma{V}=0\,,
\end{equation}
obeying the constraint
\begin{equation}
 \left\langle\ma{V}|\ma{\bar V}\right\rangle\equiv
 \bar{L}^\Lambda M_\Lambda-L^\Lambda \bar{M}_\Lambda=-i\,, \label{eq:sympcond}
\end{equation}
where \ma{K} is the K\"ahler potential.
Alternatively one can introduce the explicitly holomorphic sections of a different symplectic bundle,
\begin{equation}
 v \equiv e^{-\mathcal{K}/2}\ma{V}\equiv\left(\begin{array}{c}
						  X^\Lambda\\
						  F_\Lambda
						  \end{array}\right)\,.
\end{equation}
In appropriate symplectic frames it is possible to choose a homogeneous function of second degree $F(X)$, called prepotential, such
that $F_\Lambda=\partial_\Lambda F$.
In terms of the sections $v$ the constraint (\ref{eq:sympcond}) becomes
\begin{equation}
\label{eq:sympcond2}
 \left\langle v|\bar{v}\right\rangle\equiv\bar{X}^\Lambda F_\Lambda-X^\Lambda{\bar{F}}_\Lambda=
 -i e^{-\mathcal{K}}.
\end{equation}
The couplings of the vector fields to the scalars are determined by the $(n_V+1)\times(n_V+1)$ period matrix \ma{N}, defined by the relations
\begin{equation}
 M_\Lambda = \ma{N}_{\Lambda\Sigma}\, L^\Sigma\,, 
 \qquad D_{\bar\imath}\bar{M}_\Lambda=\ma{N}_{\Lambda\Sigma}\,D_{\bar \imath}\bar{L}^\Sigma\,.
\end{equation}
If the theory is defined in a frame in which a prepotential exists, \ma{N} can be obtained from
\begin{equation}
  \label{eq:period_matrix_prep}
  \ma{N}_{\Lambda\Sigma}=\bar{F}_{\Lambda\Sigma}
  + 2i\frac{(N_{\Lambda\Gamma}X^\Gamma)(N_{\Sigma\Delta}X^\Delta)}{X^\Omega N_{\Omega\Psi}X^\Psi}\,,
\end{equation}
where $F_{\Lambda\Sigma}=\partial_\Lambda\partial_\Sigma F$ and $N_{\Lambda\Sigma}\equiv\mathrm{Im}(F_{\Lambda\Sigma})$.
Introducing the matrix\footnote{We use the notation $R=\mathrm{Re}\,\ma N$ and
$I=\mathrm{Im}\,\ma N$.}
\begin{equation}
 \ma{M}=\left(\begin{array}{cc}
 I + 
 R  I^{-1}  R & \,\,-  R I ^{-1} \\
- I ^{-1}  R &  I ^{-1} \\
\end{array}\right),
\end{equation}
we have the important relation between the symplectic sections and their derivatives,
\begin{equation}
\frac12 (\mathcal M - i\Omega) = \Omega\bar{\mathcal V}\mathcal V\Omega + \Omega D_i\mathcal V
g^{i\bar\jmath}D_{\bar\jmath}\bar{\mathcal V}\Omega\,, \label{eq:sympid}
\end{equation}
where
\begin{equation}
\Omega = \left(\begin{array}{cc} 0 & -1 \\ 1 & 0 \end{array}\right)\,.
\end{equation}
The $4 n_H$ real hyperscalars $q^u$ parametrize a quaternionic K\"ahler manifold with metric $h_{uv}(q)$.
A quaternionic K\"ahler manifold is a $4n$-dimensional Riemannian manifold admitting a locally defined triplet 
$\vec{K}_u^{\phantom{u}v}$ of almost complex structures satisfying the quaternion relation
 \begin{equation}
   \label{eq:quaternionic_kahler_complexstruct_definition}
h^{st}K^x_{\phantom{x}us}K^y_{\phantom{y}tw}=-\delta^{xy}h_{uw}+\varepsilon^{xyz}K^z_{\phantom{z}uw}\,,
 \end{equation}
and whose Levi-Civita connection preserves $\vec{K}$ up to a rotation,
 \begin{equation}
  \label{eq:quaternionic_kahler_complexstruct_rotation}
  \nabla_w {\vec K}_u^{\phantom{u}v}+\,\vec{\omega}_w\times{\vec K}_u^{\phantom{u}v}=0\,,
 \end{equation}
where $\vec\omega\equiv \vec\omega_u (q)\, dq^u$ is the connection of the $\mathrm{SU}(2)$-bundle for which the quaternionic manifold is the base. An important property is that the $\mathrm{SU}(2)$ curvature is proportional to the
complex structures,
 \begin{equation}
  \label{eq:quaternionkahl_su2curv}
  \Omega^x\equiv\, d\omega^x+\frac12\varepsilon^{xyz}\omega^y\wedge\omega^z=-\,K^x\,.
 \end{equation}
As far as the gaugings are concerned, we shall consider only abelian symmetries of the action. 
Under abelian symmetries, the complex scalars $z^i$ transform trivially, so that we will be
effectively gauging abelian isometries of the quaternionic-K\"ahler metric $h_{uv}$.
These are generated by commuting Killing vectors $k_\Lambda^u (q)$, i.e., $[k_\Lambda,k_\Sigma]=0$.
The requirement that the quaternionic K\"ahler structure be preserved implies the existence, for each
Killing vector, of a triplet of Killing potentials, or moment maps, $P_\Lambda^x$, such that
\begin{equation}
\label{eq:mommaps}
\masf{D}_u P_\Lambda^x\equiv \partial_u P_\Lambda^x + \varepsilon^{xyz}\omega_{\phantom{y} u}^y
P_\Lambda^z = -2\Omega^x{}_{uv} k_\Lambda^v\,.
\end{equation}
One of the most important relations satisfied by the moment maps is the so-called equivariance relation.
For abelian gaugings it has the form
\begin{equation}
\frac12\epsilon^{xyz} P^x_\Lambda P^y_\Sigma - \Omega^x_{uv} k^u_\Lambda k^v_\Sigma =  0\,.
\label{eq:equivariance}
\end{equation}
The bosonic Lagrangian reads
\begin{equation}
\begin{split}
\label{eq:mainaction}
\sqrt{-g}^{-1}\!\mathscr{L} &= \frac R2 - g_{i\bar\jmath}\,\partial_{\mu}z^i\partial^{\mu}
\bar z^{\bar\jmath} - h_{uv}\hat{\partial}_{\mu} q^u\hat{\partial}^{\mu} q^v\\
   &+\frac14 I_{\Lambda\Sigma}F^{\Lambda\mu\nu}F^{\Sigma}{}_{\mu\nu} 
  + \frac14 R_{\Lambda\Sigma}F^{\Lambda\mu\nu}\star\! F^{\Sigma}{}_{\mu\nu} - V_g(z,\bar z,q)\,,
\end{split}
\end{equation}
where the scalar potential has the form
\begin{equation}
\label{eq:scal_pot}
V_g = 4h_{uv} k^u_{\Lambda} k^v_\Sigma L^\Lambda\bar{L}^\Sigma + (g^{i\bar\jmath}D_i L^{\Lambda}
D_{\bar\jmath}\bar{L}^{\Sigma} - 3 L^\Lambda\bar{L}^{\Sigma})P^x_\Lambda P^x_\Sigma\,, 
\end{equation}
the covariant derivatives acting on the hyperscalars are
\begin{equation}
\hat{\partial}_\mu q^u = \partial_\mu q^u + A^\Lambda_\mu k_\Lambda^u\,,
\end{equation}
and
\begin{equation}
I_{\Lambda\Sigma}\equiv\mathrm{Im}\,\ma{N}_{\Lambda\Sigma}\,, \qquad R_{\Lambda\Sigma}\equiv\mathrm{Re}\,\ma{N}_{\Lambda\Sigma}\,,
\qquad I^{\Lambda\Sigma} I_{\Sigma\Gamma}=\delta^\Lambda{}_\Gamma\,.
\end{equation}

\section{Hamilton-Jacobi, flow equations and magnetic gaugings}
\label{sec:flow-eqns}

In this section, we impose staticity and spherical or hyperbolic symmetry on the solutions. The resulting
equations of motion can then be derived from a one-dimensional effective action that can be written
as a sum of squares by using the Hamilton-Jacobi formalism. This will lead to first-order flow equations
in presence of both electric and magnetic gaugings.

\subsection{\label{effectiveaction}Effective action and Hamiltonian}

If we introduce the quantities
\begin{equation}
\mathcal Q^x = \langle\mathcal P^x, \mathcal Q\rangle = p^\Lambda P^x_\Lambda\,,
\qquad \mathcal W^x = \langle\mathcal P^x, \mathcal V\rangle = L^\Lambda P^x_\Lambda \,,
\end{equation}
with 
\begin{equation}
 \ma{P}^x=\left(\begin{array}{c}
                   0\\
                  P^x_\Lambda
                  \end{array}\right)\,,
                  \label{eq:electricmomentmaps}
\end{equation}
and use the quaternionic relations (\ref{eq:quaternionic_kahler_complexstruct_definition}),
(\ref{eq:quaternionkahl_su2curv}), (\ref{eq:mommaps}), the scalar potential (\ref{eq:scal_pot}) can be
rewritten in the form
\begin{equation}
\label{eq:superV}
V_g = \tilde{\mathbb G}^{AB}\mathbb D_A\mathcal W^x\mathbb D_B\bar{\mathcal W}^x -
3|\mathcal W^x|^2\,,
\end{equation}
where we defined
\begin{equation}
\tilde{\mathbb G}^{AB} = \left(\begin{array}{cc}
g^{i\bar\jmath} & 0 \\
0 & \frac13 h^{uv}\\
\end{array}\right)\,, \qquad \mathbb D_A = \left(\begin{array}{c} D_i \\ \masf{D}_u\end{array}\right)\,.
\end{equation}
The most general static metric with spherical or hyperbolic symmetry has the form
\begin{equation}
\D s^2 = - e^{2U(r)} \D t^2 + e^{-2U(r)} \D r^2 + e^{2(\psi(r) - U(r))}\D\Omega_{\kappa}^2\,,\label{eq:ansatzmet} 
\end{equation}
where $\D\Omega_{\kappa}^2=\D\theta^2+f_{\kappa}^2(\theta)\D\varphi^2$ is the metric on the two-dimensional surfaces $\Sigma=\{\mathrm{S}^2,\mathrm{H}^2\}$ of constant scalar curvature
$R=2\kappa$, with $\kappa\in\{1,-1\}$, and
\begin{equation}
f_\kappa(\theta) = \frac{1}{\sqrt{\kappa}} \sin(\sqrt{\kappa}\theta) = 
\left\{\begin{array}{c@{\quad}l} \sin\theta\, & \kappa=1\,, \\                                             
                                             \sinh\theta\, & \kappa=-1\,. \end{array}\right.
\end{equation} 
The scalar fields depend only on the radial coordinate,
\begin{equation}
z^i = z^i(r)\,, \qquad q^u = q^u(r)\,,
\label{eq:ansatzscalars}
\end{equation}
while the abelian gauge fields $A^\Lambda$ are given by
\begin{equation}
\label{eq:ansatzgaugefields}
A^\Lambda = A^\Lambda_t(r)\D t - \kappa p^\Lambda f^{\prime}_{\kappa}(\theta)\D\phi\,.
\end{equation}
Their field strengths $F^\Lambda=\D A^\Lambda$ must have the form
\begin{equation}
F^\Lambda_{tr} = e^{2(U - \psi)} I^{\Lambda\Sigma}\left(R_{\Sigma\Gamma} p^\Gamma -
e_\Sigma(r)\right)\,, \qquad F^\Lambda_{\theta\phi} = p^\Lambda f_{\kappa}(\theta)\,. \label{eq:ansatzF}
\end{equation}
The magnetic and electric charges $(p^\Lambda, e_\Lambda)$ are defined as
\begin{equation}
p^{\Lambda} = \frac1{\mbox{vol}(\Sigma_{\kappa})}\int_{\Sigma_{\kappa}} F^{\Lambda}\,, \quad   e_{\Lambda}(r) = \frac1{\mbox{vol}(\Sigma_\kappa)}\int_{\Sigma_\kappa} G_{\Lambda}\,, \quad
\mbox{vol}(\Sigma_\kappa) = \int f_\kappa(\theta)\D\theta\wedge\D\phi\,,
\label{eq:charges}
\end{equation}
where
\begin{equation}
G_{\Lambda} = -\frac2{\sqrt{-g}}\star\!\frac{\delta\mathscr{L}}{\delta F^{\Lambda}}\,.
\label{eq:Gdual}
\end{equation}
Note that the electric charges can depend on the radial coordinate. This can be easily understood,
since the running hyperscalars are electrically charged, and thus contribute to the total electric charge
inside the 2-surfaces $\Sigma_\kappa(r)$ of constant $r$ and $t$.
In fact, the Maxwell equations obtained by varying \eqref{eq:mainaction} w.r.t.~$A^\Lambda_\mu$ read
\begin{equation}
\partial_\mu(\sqrt{-g}\star\! G_\Lambda^{\phantom{\Lambda}\mu\nu})  = -2\sqrt{-g}\, h_{uv}
k^u_\Lambda\hat{\partial}^\nu q^v\,. \label{eq:max}
\end{equation}
Imposing the ansatz (\ref{eq:ansatzmet}), (\ref{eq:ansatzscalars}) and (\ref{eq:ansatzgaugefields}) on the
$t$-component, one obtains the radial variation of the electric charges,
\begin{equation}
e^\prime_{\Lambda} = -2e^{2\psi - 4U} h_{uv} k^u_\Lambda k^v_\Sigma A^\Sigma_t\,.
\label{eq:maxansatz-t}
\end{equation}
On the other hand, the magnetic charges are always constant as a consequence of the Bianchi identities
$\nabla_\nu\star F^{\Lambda\mu\nu}=0$.

The equations of motion following from (\ref{eq:mainaction}) with the ansatz (\ref{eq:ansatzmet}), (\ref{eq:ansatzscalars}) and (\ref{eq:ansatzgaugefields}) can also be obtained from the effective action
\begin{equation}
S = \int\D r L = \int\D r\left[e^{2\psi}\left(U'^2 - \psi'^2 + h_{u v} q^{\prime\,u} q^{\prime\,v} +
g_{i\bar{\jmath}} z^{\prime\,i}\bar{z}^{\prime\,\bar{\jmath}}\right) + e_\Lambda A^{\prime\,\Lambda}_t
- V\right]\,, \label{eq:redlagrangian} 
\end{equation}
where $V$ is given by
\begin{equation}
V = - e^{2(U-\psi)}V_{\text{BH}} + e^{2\psi-4U} h_{uv} k^u_\Lambda k^v_\Sigma A^\Lambda_t
A^\Sigma_t + \kappa - e^{2(\psi-U)} V_g\,, \label{eq:V1d}
\end{equation}
with $V_{\text{BH}}$ to be defined below.
In addition to the equations of motion following from (\ref{eq:redlagrangian}), one has to impose the
Hamiltonian constraint
\begin{equation}
H = L - e_\Lambda A^{\prime\,\Lambda}_t + 2V = 0\,, \label{eq:Hamcons} 
\end{equation}
the $\varphi$-component of the Maxwell equations (\ref{eq:max})\footnote{Plugging the
spherical/hyperbolic ansatz into the $\varphi$-component of the Maxwell equations, one obtains
$p^\Lambda k_\Lambda^u k_{u\Sigma}=0$, which implies (\ref{eq:maxansatz-phi}). The
$\theta$-component is trivial.},
\begin{equation}
p^\Lambda k_\Lambda^u = 0\,, \label{eq:maxansatz-phi}
\end{equation}
as well as the $r$-component
\begin{equation}
k_{\Lambda u} q'^u=0\,. \label{eq:maxansatz-r}
\end{equation}
The effective potential $V$ is determined by the scalar potential $V_g$, the charge-dependent black hole
potential $V_{\text{BH}}$, and by a contribution coming from the covariant derivatives of the hyperscalars
plus a constant term depending on the scalar curvature $\kappa$. In particular, $V_{\text{BH}}$ can be
written in the symplectically covariant form
\begin{equation}
V_{\text{BH}} = -\frac12\ma{Q}^T\ma{M}\ma{Q}\,, \qquad \ma{Q} \equiv \left(\begin{array}{c}
p^\Lambda \\ e_\Lambda \end{array}\right)\,.
\end{equation}
Notice that the effective action (\ref{eq:redlagrangian}) does not result by merely substituting the ansatz
(\ref{eq:ansatzmet}), (\ref{eq:ansatzscalars}), (\ref{eq:ansatzgaugefields}) into the general action
(\ref{eq:mainaction}). This can be seen from $V_{\text{BH}}$ in (\ref{eq:V1d}), that does not arise by
rewriting the kinetic terms of the gauge fields. In fact it is easy to see that the gauge fields enter the
equations of motion of the whole system via their stress-energy tensor, whose components are expressed
in terms of $V_{\text{BH}}$ \cite{Ferrara:1996dd,Bellucci:2008cb,Chimento:2015rra}.

In this sense, the presence of the term $e_\Lambda A_t^{\prime\,\Lambda}$ is necessary for having the right dynamics of the variables $e_\Lambda$ and $A_t^\Lambda$. Indeed, varying the effective action
(\ref{eq:redlagrangian}) w.r.t.~$A_t^\Lambda$, one obtains exactly (\ref{eq:maxansatz-t}). Variation
w.r.t.~$e_\Lambda$ yields
\begin{equation}
A^{\prime\,\Lambda}_t = - e^{2(U-\psi)} I^{\Lambda\Sigma}(R_{\Sigma\Gamma}p^\Gamma - e_\Sigma(r))\,,
\label{eq:max1}
\end{equation}
which is exactly the expression \eqref{eq:ansatzF} for the $(t,r)$-component of $F^{\Lambda\mu\nu}$. 

Introducing
\begin{equation}
\ma{H}_{\Lambda\Sigma} = k^u_\Lambda h_{uv} k^v_\Sigma\,, \label{eq:matriceH}
\end{equation}
(\ref{eq:maxansatz-t}) becomes 
\begin{equation}
e^{\prime}_{\Lambda} = - 2 e^{2\psi-4U}\ma{H}_{\Lambda\Sigma}A^\Sigma_t\,,
\label{eq:max2}
\end{equation}
which allows to express $A^\Sigma_t$ in terms of the other fields as follows. Since
$\ma{H}_{\Lambda\Sigma}$ is real and symmetric, there exists a matrix $O\in\text{O}(n_V+1)$ such that
\begin{equation}
\ma{H}_{\Lambda\Sigma} = (O^TDO)_{\Lambda\Sigma} = O^\Omega{}_\Lambda O^\Gamma{}_\Sigma
D_{\Omega\Gamma}\,,
\end{equation}
with $D$ diagonal. Without loss of generality, suppose that the first $n$ eigenvalues of $D$ are
nonvanishing ($0\le n\le n_V+1$), while the remaining ones are zero. Let hatted indices
$\hat\Lambda,\hat\Sigma,\ldots$ range from $0$ to $n-1$, and define
\begin{equation}
{\hat A}^\Gamma_t \equiv O^\Gamma{}_\Sigma A^\Sigma_t\,.
\end{equation}
\eqref{eq:max2} yields then
\begin{equation}
O_{\hat\Psi}{}^\Lambda e^{\prime}_\Lambda = -2 e^{2\psi-4U}D_{\hat\Psi\hat\Gamma}
{\hat A}^{\hat\Gamma}_t\,, \label{eq:Oeprime}
\end{equation}
where indices are raised and lowered with the flat metric, i.e.,
$O_\Psi{}^\Lambda\equiv\delta_{\Psi\Omega}\delta^{\Lambda\Gamma}O^\Omega{}_\Gamma$.
We also get
\begin{equation}
O^{\Psi\Lambda} e^{\prime}_{\Lambda} = 0 \quad \text{for} \quad \Psi\ge n\,. \label{eq:Oeprime2}
\end{equation}
\eqref{eq:Oeprime} gives
\begin{equation}
{\hat A}^{\hat\Lambda}_t = -\frac12 e^{4U-2\psi} {(D^{-1})}^{\hat\Lambda\hat\Psi}
O_{\hat\Psi}{}^\Lambda e^{\prime}_{\Lambda}\,. \label{eq:A}
\end{equation}
Using these relations in the effective action (\ref{eq:redlagrangian}) to eliminate $A^\Sigma_t$, one obtains
\begin{equation}
S = \int\D r\left[e^{2\psi}(U^{\prime\,2} - \psi^{\prime\,2} + h_{u v} q^{\prime\,u} q^{\prime\,v} +
g_{i\bar{\jmath}} z^{\prime\,i}\bar{z}^{\prime\,\bar{\jmath}} + \frac14 e^{4(U-\psi)}
\ma{H}^{\Lambda\Sigma} e'_{\Lambda} e'_{\Sigma}) - \tilde{V}\right]\,, \label{eq:Seff}
\end{equation}
where we defined the effective potential
\begin{equation}
\tilde{V} = -e^{2(U-\psi)} V_{\text{BH}} + \kappa - e^{2(\psi-U)} V_{g}\,, \label{eq:effpot}
\end{equation}
as well as
\begin{equation}
\ma{H}^{\Lambda\Sigma} \equiv O_{\hat\Lambda}{}^\Lambda (D^{-1})^{\hat\Lambda\hat\Sigma}
O_{\hat\Sigma}{}^\Sigma\,. \label{eq:H-upper}
\end{equation}
Note that, unless $n=n_V+1$, $\ma{H}^{\Lambda\Sigma}$ is not the inverse of
$\ma{H}_{\Lambda\Sigma}$ (which is not invertible), but we have the weaker relation
\begin{equation}
\ma{H}^{\Lambda\Gamma}\ma{H}_{\Lambda\Sigma}\ma{H}_{\Gamma\Omega} = \ma{H}_{\Sigma\Omega}\,,
\label{eq:HHH}
\end{equation}
that will be used below to square the action.

One can then rewrite the constraint \eqref{eq:Hamcons} in terms of the effective Hamiltonian
\begin{equation}
H = \frac14 e^{-2\psi} p_U^2 - \frac14 e^{-2\psi} p_{\psi}^2+ \frac14 e^{-2\psi}h^{u v} p_{q^u} p_{q^v}
+ e^{-2\psi}g^{i\bar{\jmath}} p_{z^i} p_{\bar{z}^{\bar{\jmath}}} + e^{4(U-\psi)}\ma{H}^{\Lambda\Sigma}
p_{e_{\Lambda}} p_{e_{\Sigma}} + \tilde{V}\,,
\label{eq:Heff}
\end{equation}
where the canonical momenta $p_U$, $p_\psi$, $p_{q^u}$, $p_{z^i}$, $p_{\bar z^{\bar\jmath}}$
and $p_{e_\Lambda}$ are defined in the usual way.
The effective action (\ref{eq:Seff}), together with the relations \eqref{eq:Hamcons}, \eqref{eq:maxansatz-phi},
\eqref{eq:maxansatz-r}, reproduces the complete set of equations of motion for the spherical/hyperbolic
ansatz (\ref{eq:ansatzmet}), (\ref{eq:ansatzscalars}) and (\ref{eq:ansatzgaugefields}).

\subsection{\label{Flow Equations}Flow equations with electric gaugings}

Inspired by \cite{Dall'Agata:2010gj}, we aim to find first-order flow equations for the effective action
(\ref{eq:Seff}) with gauged abelian symmetries generated by the electric Killing vectors $k_\Lambda^u$,
using the Hamilton-Jacobi approach \cite{Andrianopoli:2009je,Trigiante:2012eb}. In particular, introducing Hamilton's charcteristic function associated to (\ref{eq:Seff}), one can write the action as a sum of squares
from which one can derive the flow equations\footnote{These are of course equivalent to the usual
first-order equations in the Hamilton-Jacobi formalism, but we find it convenient to explicitely
show the squaring of the action.}.

The particular form of the scalar potential (\ref{eq:superV}) gives a first hint on how a putative
Hamilton-Jacobi function may look like. Indeed, if we define
\begin{equation}
\ma L = \ma Q^x\ma W^x = p^\Lambda P^x_\Lambda L^\Sigma P^x_\Sigma\,,
\end{equation}
and require spherical/hyperbolic invariance, we can rewrite the scalar potential (\ref{eq:superV}) in a
way analogous to \cite{Dall'Agata:2010gj}. Namely, using (\ref{eq:maxansatz-phi}), the quaternionic
relations (\ref{eq:quaternionic_kahler_complexstruct_definition}), (\ref{eq:quaternionkahl_su2curv}),
(\ref{eq:mommaps}), (\ref{eq:equivariance}) and imposing\footnote{Notice that
$\partial_\mu(\ma Q^x\ma Q^x)=\partial_u(\ma Q^x\ma Q^x)\partial_\mu q^u$, and
$\partial_u(\ma Q^x\ma Q^x)=\masf{D}_u(\ma Q^x\ma Q^x)=2\ma{Q}^x\masf{D}_u\ma{Q}^x$. Using
the definition of $\ma Q^x$ together with \eqref{eq:mommaps}, this is equal to
$-4\ma{Q}^x p^\Lambda\Omega^x{}_{uv}k^v_\Lambda$, which vanishes by virtue
of \eqref{eq:maxansatz-phi}. $\ma Q^x\ma Q^x$ is thus a constant of motion, that we choose to be one.}
\begin{equation}
\ma Q^x\ma Q^x =1\,,
\label{eq:quantization}
\end{equation}
one can show that the scalar potential (\ref{eq:superV}) can be expressed in terms of the superpotential
$\ma L$ as
\begin{equation}
V_g = \mathbb G^{AB}\mathbb D_A\ma L\,\mathbb D_B\bar{\ma L} - 3|\ma L|^2\,, \label{eq:superV1}
\end{equation}
where
\begin{equation}
\mathbb G^{AB} = \left(\begin{array}{cc}
g^{i\bar\jmath} & 0 \\ 0 & h^{uv} \end{array}\right)\,, \qquad
\mathbb D_A = \left(\begin{array}{l} D_i \\ \masf{D}_u \end{array}\right)\,.
\label{eq:globalmetric}
\end{equation}
However, the effective potential \eqref{eq:effpot} contains not only $V_g$, and thus Hamilton's
characteristic function $W$ (that solves the `time' (i.e., $r$)-independent HJ equation) must contain
also other contributions in addition to $\ma L$. This happens also in the case without hypermultiplets
and $\text{U}(1)$ Fayet-Iliopoulos gauging \cite{Dall'Agata:2010gj,Klemm:2012vm}.
When there are also running hyperscalars, the general structure of the effective action remains essentially
the same except for the presence of some new kinetic terms. The main difference is the form of the scalar
potential $V_g$,  which is now governed by the superpotential $\ma L$, that depends on the
tri-holomorphic moment maps. Guided by these observations, and following \cite{Dall'Agata:2010gj}, we
introduce the real function
\begin{equation}
W = e^U |\mathcal Z  + i\kappa e^{2\psi - 2U} \ma L|\,,
\label{eq:superpotential}
\end{equation}
and a phase $\alpha$ defined by
\begin{equation}
e^{2i \alpha} = \frac{\mathcal Z  + i\kappa e^{2(\psi - U)}\mathcal L}{\bar{\mathcal Z} - i\kappa
e^{2(\psi - U)}\bar{\mathcal L}}\,, \qquad \mbox{or} \qquad \mathrm{Im}(e^{-i\alpha}\mathcal Z) =
- \kappa e^{2(\psi - U)}\mathrm{Re}(e^{-i\alpha}\mathcal L)\,, \label{eq:defal}
\end{equation}
where $\ma Z= \langle\ma Q,\ma V\rangle$ is the central charge. Defining `tilded' variables by
$\tilde{\ma X}=e^{-i\alpha}\ma X$ etc., we can rewrite $W$ as
\begin{equation}
W = e^U\mathrm{Re}\tilde{\ma Z} - \kappa e^{2\psi - U}\mathrm{Im}\tilde{\ma L}\,.
\label{eq:Wnd}
\end{equation}
Using (\ref{eq:sympid}), (\ref{eq:mommaps}) and (\ref{eq:quantization}), it is possible to shew that
\begin{eqnarray}
&& e^{-2\psi}\left((\partial_U W)^2 - (\partial_\psi W)^2 + 4 g^{i\bar{\jmath}}\partial_i W
\partial_{\bar{\jmath}} W +  h^{uv}\partial_u W\partial_v W + 4 e^{4(\psi - U)}\ma H_{\Lambda\Sigma}
\partial_ {e_\Lambda} W\partial_{e_\Sigma} W\right) \nonumber \\
&& \qquad -e^{2(\psi - U)}  V_g -  e ^{2(U - \psi)} V_{\text{BH}} + \kappa=0\,, \label{eq:sqpot}
\end{eqnarray}
or, in other words, that $2W$ solves the Hamilton-Jacobi equation associated to the Hamiltonian (\ref{eq:Heff}) with zero energy. By virtue of \eqref{eq:sqpot}, up to a total derivative, the action
(\ref{eq:Seff}) can be written as
\begin{equation}
\begin{split}
S &= \int\D r \Bigl[ e^{2\psi} \bigl(U^\prime + e^{-2 \psi }\partial_U W\bigl)^2 
- e^{2\psi} \bigl(\psi^\prime - e^{-2\psi}\partial_\psi W\bigl)^2 + \\
&e^{2\psi }g_{i\bar\jmath}\bigl(z^{\prime\,i} + 2e^ {-2\psi} g^{i\bar{k}}\partial_{\bar{k}} W\bigl) 
\bigl(\bar{z}^{\prime\,\bar{\jmath}} + 2e^ {-2\psi} g^{\bar{\jmath}\,l} \partial_l W\bigl) + \\
& e^{2\psi} h_{uv} \bigl(q^{\prime\,u} + e^{-2\psi} h^{us}\partial_s W\bigl)
\bigl(q^{\prime\,v} + e^{-2\psi} h^{vt}\partial_t W\bigl) + \\
&\frac14 e^{4U - 2\psi}\ma H^{\Lambda\Gamma}\bigl(e^\prime_\Lambda + 4e^{2\psi- 4U}
\ma H_{\Lambda\Sigma} \partial_{e_{\Sigma}} W\bigl)\bigl(e^\prime_{\Gamma} + 4e^{2\psi- 4U}
\ma H_{\Gamma\Omega}\partial_{e_{\Omega}} W\bigl)\Bigl]\,,
\end{split} \label{eq:BPSaction}
\end{equation}
where we used also \eqref{eq:HHH} and the fact that \eqref{eq:Oeprime2} implies
\begin{equation}
\ma H^{\Lambda\Gamma}\ma H_{\Lambda\Sigma}\partial_{e_\Sigma}W e^{\prime}_\Gamma
= \partial_{e_\Gamma} W e^{\prime}_\Gamma\,.
\end{equation}
The BPS-rewriting \eqref{eq:BPSaction} guarantees that the solutions of the first-order
equations obtained by setting each quadratic term to zero do indeed extremize the action.
If one explicitly computes the derivatives of $W$,
these first-order flow equations become
\begin{equation}
\begin{split}
&U^\prime = -e^{U-2\psi}\mathrm{Re}\tilde{\mathcal Z} - \kappa e^{-U}\mathrm{Im}\tilde{\mathcal L}\,, \\
&\psi^\prime = -2\kappa e^{-U}\mathrm{Im}\tilde{\mathcal L}\,, \\
&z^{\prime\,i} = -e^{i\alpha} g^{i\bar{\jmath}}\left(e^{U-2\psi} D_{\bar\jmath}\bar{\mathcal Z} - i\kappa
e^{-U} D_{\bar\jmath}\bar{\mathcal L} \right)\,, \\
&q^{\prime\,u} = \kappa e^{-U} h^{uv}\mathrm{Im}(e^{-i\alpha}\partial_v\mathcal L)\,, \\
&e^\prime_\Lambda = -4 e^{2\psi-3U}\ma H_{\Lambda\Sigma}\mathrm{Re}\tilde{L}^\Sigma\,.
\end{split} \label{eq:flow-equ-electr}
\end{equation}
These relations, plus the constraints that we had to impose, are equivalent to those obtained in
\cite{Halmagyi:2013sla}\footnote{\eqref{eq:flow-equ-electr} corrects some sign errors in appendix B of
\cite{Halmagyi:2013sla}.} from the Killing spinor equations. To see this, note that comparing
the expression for $e^{\prime}_\Lambda$ in \eqref{eq:flow-equ-electr} with (\ref{eq:max2}) yields
the additional condition
\begin{equation}
2e^U\ma H_{\Lambda\Sigma}\mathrm{Re}\tilde{L}^\Sigma\ = \ma{H}_{\Lambda\Sigma}
A^\Sigma_t\,, \label{eq:ReL-A}
\end{equation}
which is just (B.44) of \cite{Halmagyi:2013sla} contracted with $h_{uv}k^v_\Sigma$. To be precise,
\eqref{eq:ReL-A} is equivalent to
\begin{equation}
2 e^U k^u_\Lambda\mathrm{Re}\tilde{L}^\Lambda = k^u_\Lambda A^\Lambda_t + m^u\,,
\end{equation}
where $m^u$ must satisfy $k^v_\Sigma h_{uv}m^u = 0\,\,\forall\,\Sigma$. If $n^\Sigma$ is an
eigenvector of $\ma H_{\Lambda\Sigma}$ with zero eigenvalue, i.e.,
$\ma H_{\Lambda\Sigma}n^\Sigma=0$, then we can take the linear combination
$m^u=k^u_\Lambda n^\Lambda$. (B.44) of \cite{Halmagyi:2013sla} has $m^u=0$, and is thus
slightly stronger than \eqref{eq:ReL-A}. Notice also that the number of independent constraints
coming from \eqref{eq:ReL-A} is equal to $n$, where $n$ denotes the number of nonvanishing
eigenvalues of $\ma H_{\Lambda\Sigma}$. This becomes evident by casting \eqref{eq:ReL-A} into
the form
\begin{equation}
2 e^U D_{\Omega\Gamma}O^\Gamma{}_\Sigma\mathrm{Re}\tilde{L}^\Sigma =
D_{\Omega\Gamma}\hat A^\Gamma_t\,.
\end{equation}
The auxiliary field $\alpha$ is related to the phase of the Killing spinor associated to the BPS solution,
as was shown for the case without hypers and $\text{U}(1)$ Fayet-Iliopoulos gauging
in \cite{Dall'Agata:2010gj}\footnote{Without hypermultiplets and for $\text{U}(1)$ FI gauging,
one can always choose $P^1_\Lambda = P^2_\Lambda = 0$, $P^3_\Lambda \equiv \mathcal G$ for the
moment maps by a global $\text{SU}(2)$ rotation (which is a symmetry of the theory). The condition
\eqref{eq:quantization} becomes then $\ma Q^3=\langle\mathcal G,\mathcal Q\rangle=-\kappa$, and
the function $W$ boils down to equ.~(2.40) of \cite{Dall'Agata:2010gj} for $\kappa=1$.}, and for
the case including hypermultiplets in \cite{Halmagyi:2013sla}.

Finally, since the eqns.~\eqref{eq:flow-equ-electr} describe extremal configurations, there exists an
additional constant of motion $\mathbb Q$ \cite{Trigiante:2012eb} such that
\begin{equation}
\frac {\D\mathbb Q}{\D r} = H = 0\,.
\end{equation}
Using the first order equations for $U$ and $\psi$, one gets from \eqref{eq:sqpot}
\begin{equation}
\mathbb Q = e^{2 \psi }(U' - \psi') + W\,.
\end{equation}

\subsection{\label{magnetic Gaugings}Magnetic gaugings and symplectic covariance}

The most natural way to extend the results of the last section is to consider also magnetic gauge fields
$A_{\Lambda\mu}$. This implies the inclusion of magnetic Killing vectors $k^{\Lambda u}$ and magnetic
moment maps $P^{x\Lambda}$.
This formulation of gauged supergravity is typically expressed in terms of the embedding tensor
$\Theta^a_M=(\Theta^a_\Lambda,\Theta^{a\Lambda} )^T$\footnote{In this section we explicitly
introduce the indices ($M,N,\ldots$) in the fundamental representation of $\text{Sp}(2n_V+2,\mathbb R)$
for clarity \cite{deWit:2005ub,Samtleben:2008pe}.}, and the main consequence is the restoration of
symplectic covariance of the theory \cite{deWit:2005ub,Samtleben:2008pe}.

In this context, one introduces the symplectic vectors
\begin{equation}
\ma{A}_{\mu} = \left(\begin{array}{c}
A_{\mu}^{\Lambda} \\ A_{\Lambda\mu} \end{array}\right)\,, \qquad
\mathcal K^u = \left(\begin{array}{c}
k^{\Lambda u} \\ k_\Lambda^u \end{array}\right)\,, \qquad 
\mathcal P^x = \left(\begin{array}{c}
P^{x\Lambda} \\ P^x_\Lambda \end{array}\right)\,, \label{eq:magneticgaugings}
\end{equation}
where the magnetic quantities $k^{\Lambda u}$ and $P^{x\Lambda}$ obey the relations introduced
in section \ref{Setup}. As was shown in \cite{Erbin:2014hsa}, the locality constraint
$\langle\Theta^a,\Theta^b\rangle=0$, namely the possibility to rotate any gauging to a frame with a
purely electric one, implies also
\begin{equation}
\langle\ma K^u,\ma P^x\rangle = 0\,.
\label{eq:cons}
\end{equation}
In presence of magnetic gaugings, the general action (\ref{eq:mainaction}) is modified in a nontrivial way
by some topological terms \cite{deWit:2005ub}. The consistency of the theory requires the introduction of
the auxiliary 2-forms $B_a=\frac12 B_{a\mu\nu}\D x^\mu\wedge\D x^\nu$ that do not change
the number of degrees of freedom. The action has the form \cite{deWit:2005ub,Samtleben:2008pe}
\begin{equation}
\begin{split}
\sqrt{-g}^{-1}\!\mathscr{L} &= \frac R2 - g_{i\bar\jmath}\,\partial_\mu z^i\partial^\mu
\bar z^{\bar\jmath} - h_{uv}\hat{\partial}_\mu q^u\hat{\partial}^\mu q^v + \frac14 I_{\Lambda\Sigma}
H^{\Lambda\mu\nu} H^{\Sigma}{}_{\mu\nu} + \\
& \frac14 R_{\Lambda\Sigma} H^{\Lambda\mu\nu}\star\! H^{\Sigma}{}_{\mu\nu} -
\frac{\epsilon^{\mu\nu\rho\sigma}}{4\sqrt{-g}}\Theta^{a\Lambda} B_{a\mu\nu}\partial_\rho
A_{\Lambda\sigma} + \\
& \frac1{32\sqrt{-g}}\Theta^{\Lambda a}\Theta_\Lambda^b\epsilon^{\mu\nu\rho\sigma} B_{a\mu\nu}
B_{b\rho\sigma} - V_g\,, \label{actionmag}
\end{split}
\end{equation}
where the modified field strength $H^\Lambda{}_{\mu\nu}=F^\Lambda{}_{\mu\nu}+\frac12
\Theta^{\Lambda a} B_{a\mu\nu}$ was introduced. The co\-var\-iant derivatives of the hyperscalars and the
scalar potential read respectively \cite{Samtleben:2008pe,Michelson:1996pn,deWit:2005ub}
\begin{equation}
\hat{\partial}_\mu q^u = \partial_\mu q^u - A_\mu^\Lambda \Theta_\Lambda^a k_a^u -
A_{\Lambda\mu}\Theta^{\Lambda a} k_a^u \equiv \partial_\mu q^u - \langle\ma A_\mu,\ma K^u\rangle\,,
\label{eq:covmag}
\end{equation}
\begin{equation}
V_g = 4 h_{uv} \langle\mathcal K^u,\mathcal V\rangle\langle\mathcal K^v,\bar{\mathcal V}\rangle
+ g^{i\bar{\jmath}}\langle\mathcal P^x, D_i\mathcal V\rangle\langle\mathcal P^x,\bar {D}_{\bar{\jmath}}
\bar{\mathcal V}\rangle - 3\langle\mathcal P^x,\mathcal V\rangle\langle\mathcal P^x,
\bar{\mathcal V}\rangle\,. \label{eq:Vg-sympl-cov}
\end{equation}
Note that it is also possible to generate \eqref{eq:Vg-sympl-cov} from \eqref{eq:superV} by a
symplectic rotation.

The equations of motion for $A_{\Lambda\mu}$, $A_\mu^\Lambda$ and $B_{a\mu\nu}$ following from
(\ref{actionmag}) are
\begin{equation}
\begin{split}
&\frac14 \epsilon^{\mu\nu\rho\sigma}\partial_\mu B_{a\nu\rho}\Theta^{\Lambda a} = -2\sqrt{-g}
h_{uv}\Theta^{\Lambda a} k_a^u\hat{\partial}^\sigma q^v\,, \\
&G_{\Lambda\mu\nu}\Theta^{\Lambda a} = \Theta^{\Lambda a}(F_{\Lambda\mu\nu} - \frac12
\Theta_\Lambda^b B_{b\mu\nu})\,, \\
&\partial_\mu\Bigl(\sqrt{-g} I_{\Lambda\Sigma} H^{\Sigma\mu\nu} + \frac12
\epsilon^{\mu\nu\rho\sigma} R_{\Lambda\Sigma} H^\Sigma{}_{\rho\sigma}\Bigl) = 2\sqrt{-g} h_{uv}
\Theta^a _\Lambda k^u_a\hat{\partial}^\nu q^v\,,
\label{eq:eommag1}
\end{split}
 \end{equation}
where $G_{\Lambda\mu\nu}$ is defined by \eqref{eq:Gdual}. The eqns.~\eqref{eq:eommag1} can be
rewritten in a completely symplectically covariant form as
\begin{equation}
\frac12\epsilon^{\mu\nu\rho\sigma}\partial_\nu G_{\rho\sigma}^M = \Omega^{MN} J^\mu_N\,, \qquad
\Theta^{a M} (H-G)_M = 0\,, 
\end{equation}
where
\begin{equation}
H^M_{\mu\nu} = F^M_{\mu\nu} + \frac12\Omega^{MN}\Theta^a_N B_{a\mu\nu}\,, \qquad
G_{\mu\nu}^M = (H^\Lambda{}_{\mu\nu},G_{\Lambda\mu\nu})\,,
\end{equation}
and $J^\mu _M$ are the currents coming from the coupling to the matter. This result is exactly what one
expects in presence of magnetic and electric sources for the Maxwell fields. In this context, it is clear that
both the electric and magnetic charges will depend on the radial coordinate, once we impose spherical or 
hyperbolic symmetry.

The latter implies the following form for the electric and magnetic gauge fields and the 2-forms $B_a$,
\begin{equation}
A^\Lambda = A_t^\Lambda\D t- \kappa p^\Lambda f^\prime_\kappa(\theta)\D\phi\,, \qquad
A_\Lambda = A_{\Lambda t}\D t - \kappa e_\Lambda f^\prime_\kappa(\theta)\D\phi\,,
\end{equation}
\begin{equation}
B^\Lambda = 2\kappa p^{\prime\,\Lambda} f^\prime_\kappa(\theta)\D r\wedge\D\phi\,, \qquad 
B_\Lambda = -2\kappa e^{\prime}_\Lambda f^\prime_\kappa(\theta)\D r\wedge\D\phi\,,
\end{equation}
which implies for the field strengths
\begin{equation}
H^\Lambda{}_{tr} = e^{2(U - \psi)} I^{\Lambda\Sigma}(R_{\Sigma\Gamma} p^\Gamma - e_\Sigma)\,,
\qquad H^\Lambda{}_{\theta\phi} = p^\Lambda f_\kappa(\theta)\,,
\end{equation}
\begin{equation}
G_{\Lambda tr} = e^{2(U - \psi)}\left(I_{\Lambda\Sigma} p^\Sigma + R_{\Lambda\Gamma}
I^{\Gamma\Omega} R_{\Omega\Sigma} p^\Sigma - R_{\Lambda\Gamma} I^{\Gamma\Omega}
e_\Omega\right)\,, \qquad G_{\Lambda\theta\phi} = e_\Lambda f_\kappa(\theta)\,.
\label{eq:ansatzH}
\end{equation}
Introducing the symplectic matrix
\begin{equation}
\ma H = (\mathcal K^u)^T h_{uv}\mathcal K^v\,,
\end{equation}
and plugging the above ansatz into (\ref{eq:eommag1}), one obtains
\begin{equation}
\mathcal A^\prime_t = - e^{2(U - \psi)}\Omega\mathcal M\mathcal Q\,, \qquad
\mathcal Q^\prime = -2 e^{2\psi - 4U}\mathcal H\Omega\mathcal A_t\,, \label{eq:maxsim} 
\end{equation}
where the constraints
\begin{equation}
\ma H\Omega\ma Q = 0\,, \qquad \mathcal K_u q^{\prime u} = 0
\label{eq:symcon}
\end{equation}
have been imposed. It is worthwhile to note that the first equation of \eqref{eq:symcon} permits the
rewriting of $V_g$ as in (\ref{eq:superV1}) starting from (\ref{eq:superV}), namely
\begin{equation}
V_g = \mathbb G^{AB}\mathbb D_A\ma L\mathbb D_B\bar{\ma L} - 3|\ma L|^2\,, \qquad
\ma L = \ma Q^x\ma W^x = \langle\ma Q^x\ma P^x,\ma V\rangle\,.
\end{equation}

Following the same procedure used previously for purely electric gaugings, one finds the effective
action that generalizes (\ref{eq:Seff}),
\begin{equation}
\begin{split}
&S = \int\D r\left[e^{2\psi}(U^{\prime\,2} - \psi^{\prime\,2} + h_{u v} q^{\prime\,u} q^{\prime \, v} +
g_{i\bar{\jmath}}\,z^{\prime\, i}\bar{z}^{\prime\,\bar{\jmath}} + \frac14 e^{4(U - \psi)}
{\ma Q^\prime}^T\ma{H}^{-1}\ma Q^\prime) - \tilde{V}\right]\,, \\
&\tilde{V} = -e^{ 2(U - \psi)} V_{\text{BH}}  + \kappa - e^{2(\psi - U)} V_{g}\,,
\label{eq:Seffmag}
\end{split}
\end{equation}
where, in a slight abuse of notation, $\ma H^{-1}$ denotes the symplectically covariant generalization
of the matrix $\ma H^{\Lambda\Sigma}$ defined by \eqref{eq:H-upper}. (Note that one has not
necessarily $\ma H^{-1}\ma H=\mathbb{I}$, cf.~the discussion in section \ref{effectiveaction}, but
$\ma H^{-1}$ in \eqref{eq:Seffmag} can be defined in a way similar to \eqref{eq:H-upper}).

Introducing the function $W$ and the phase $\alpha$ as in (\ref{eq:superpotential}) and (\ref{eq:defal}),
with the obvious symplectic generalization of $\mathcal L$, it is straightforward to shew that $W$
satisfies the Hamilton-Jacobi equation for the action (\ref{eq:Seffmag}),
\begin{eqnarray}
&& e^{-2\psi}\left((\partial_U W)^2 - (\partial_\psi W)^2 + 4 g^{i\bar{\jmath}}\partial_i W
\partial_{\bar{\jmath}} W +  h^{uv}\partial_u W \partial_v W + 4 e^{4(\psi - U)}({\partial_ {\ma Q}
W})^T\ma H\partial_{\ma Q }W\right) \nonumber \\
&& \qquad -e^{2(\psi - U)}  V_g - e^{2(U - \psi)} V_{\text{BH}} + \kappa = 0\,, \label{eq:sqpotmag}
\end{eqnarray}
provided the charge-quantization condition \eqref{eq:quantization} holds, with
$\mathcal Q^x = \langle\mathcal P^x, \mathcal Q\rangle$. Using \eqref{eq:sqpotmag} as well as
\eqref{eq:cons} and discarding total derivatives, the action (\ref{eq:Seffmag}) can be cast into the form
\begin{equation}
\begin{split}
S &= \int\D r\Bigl[e^{2\psi}\bigl(U^\prime + e^{-2\psi }\partial_U W\bigl)^2
- e^{2\psi}\bigl(\psi^\prime - e^{-2\psi}\partial_\psi W\bigl)^2 + \\
&e^{2\psi} g_{i\bar{\jmath}}\bigl(z^{\prime\,i} + 2 e^ {-2\psi} g^{i\bar{k}}\partial_{\bar{k}} W\bigl)
\bigl(\bar{z}^{\prime\,\bar{\jmath}} + 2 e^ {-2\psi} g^{\bar{\jmath}l}\partial_l W\bigl) + \\
& e^{2\psi} h_{uv}\bigl(q^{\prime\,u} + e^{-2\psi} h^{us}\partial_s W\bigl)
\bigl(q^{\prime\,v} + e^{-2\psi} h^{vt}\partial_t W\bigl) + \\
&\frac14 e^{4U - 2\psi}\bigl(\ma Q^\prime + 4 e^{2\psi- 4U}\ma H\partial_{\ma Q} W\bigl)^T
\ma H^{-1}\bigl(\ma Q^\prime + 4 e^{2\psi- 4U}\ma H\partial_{\ma Q} W\bigl)\Bigl]\,.
\label{eq:magsquaring}
\end{split}
\end{equation}
All first-order equations following from \eqref{eq:magsquaring} except the one for $z^i$ are 
symplectically covariant. Computing explicitely $\partial_{\bar k}W$, the latter reads
\begin{equation}
z^{\prime\,i} = - e^{i\alpha} g^{i\bar{\jmath}}\left(e^{U-2\psi} D_{\bar\jmath}\bar{\mathcal Z}
- i\kappa e^{-U} D_{\bar\jmath}\bar{\mathcal L}\right)\,.
\end{equation}
Contracting this with $D_i\ma V$ and using \eqref{eq:sympid}, one obtains a symplectically
covariant equation for the section $\ma V$,
\begin{eqnarray}
\mathcal V^\prime + i A_r\mathcal V &=& e^{i\alpha} e^{U - 2\psi}\left(-\frac12\Omega\mathcal M
\mathcal Q - \frac i2\mathcal Q + \bar{\mathcal V}\mathcal Z\right) \nonumber \\
&& - i\kappa e ^{i\alpha} e^{-U}\left(-\frac12\Omega\mathcal M\mathcal P^x\mathcal Q ^x -
\frac i2\mathcal P^x\mathcal Q^x + \bar{\mathcal V}\mathcal L\right)\,,
\end{eqnarray}
where $A_r=\mathrm{Im}(z^{\prime\,i}\partial_i\mathcal K)$ is the $\text{U}(1)$ K\"ahler connection.
Calculating the remaining derivatives of $W$, the first-order flow equations become
\begin{equation}
\begin{split}
&U^\prime = -e^{U-2\psi}\mathrm{Re}\tilde{\mathcal Z} - \kappa e^{-U}\mathrm{Im}\tilde{\mathcal L}\,, \\
&\psi^\prime = -2\kappa e^{-U}\mathrm{Im}\tilde{\mathcal L}\,, \\
&q^{\prime\,u} = \kappa e^{-U} h^{uv}\mathrm{Im}(e^{-i\alpha}\partial_v \mathcal L)\,, \\
&\ma Q^\prime = -4 e^{2\psi - 3U}\mathcal H\Omega\mathrm{Re}\tilde{\mathcal V}\,, \\
&\mathcal V^\prime  = e^{i\alpha} e^{U - 2\psi}\left(-\frac12\Omega\mathcal M\mathcal Q -
\frac i2\mathcal Q + \bar{\mathcal V}\mathcal Z\right) \\
&- i\kappa e^{i\alpha} e^{-U}\left(-\frac12\Omega\mathcal M\mathcal P^x\mathcal Q^x -
\frac i2\mathcal P^x\mathcal Q^x + \bar{\mathcal V}\mathcal L\right) - i A_r\mathcal V\,.
\label{eq:flowmag}
\end{split}
\end{equation}
These equations have a more useful form if one consider the phase $\alpha$ as a dynamical variable. 
Introducing the quantity $\mathcal S=\mathcal Z + i\kappa e^{2(\psi - U)}\mathcal L$, the relations (\ref{eq:defal}) and (\ref{eq:Wnd}) can be rewritten as
\begin{equation}
e^{2i\alpha} = \frac{\mathcal S}{\bar{\mathcal S}}\,, \qquad \mathrm{Im}(e^{-i\alpha}\mathcal S) = 0\,,
\qquad W = e^U\mathrm{Re}(e^{-i\alpha}\mathcal S)\,, \qquad W^2 = e^{2U}\mathcal S\bar{\mathcal S}\,.
\end{equation} 
One has thus
\begin{equation}
\alpha^\prime = \frac{\mathrm{Im}(e^{-i\alpha}\mathcal S^\prime)}{e^{-U} W}\,, \qquad \mathcal S^\prime
= U^\prime\partial_U\mathcal S + \psi^\prime\partial_\psi\mathcal S + \mathcal V^\prime
\partial_{\mathcal V}\mathcal S + q^{\prime\,u}\partial_u\mathcal S + \mathcal Q^{\prime\,T}
\partial_{\mathcal Q}\mathcal S\,.
\end{equation}
Inserting (\ref{eq:flowmag}) and the derivatives of $\ma S$ in this last expression, one gets
\begin{equation}
\alpha' + A_r = 2\kappa e^{-U}\mathrm{Re}(e^{-i\alpha}\mathcal L)\,.
\end{equation}
Finally, plugging the equation for $U$ into the expression of $\mathrm{Im}\tilde{\ma V}^\prime$, one
can write the first-order flow equations in the form
\begin{eqnarray}
&& 2 e^{2\psi}\left(e^{-U}\mathrm{Im}(e^{-i\alpha}\mathcal{V})\right)^{\prime} - \kappa e^{2(\psi -
U)}\Omega\mathcal{M}\ma Q^x\mathcal{P}^x + 4 e^{2\psi-U}(\alpha^{\prime} + A_r)\mathrm{Re}
(e^{-i\alpha}\mathcal{V}) + \mathcal{Q} = 0\,, \nonumber \\
&& \psi^{\prime} = -2\kappa e^{-U}\mathrm{Im}(e^{-i\alpha}\mathcal{L})\,, \nonumber \\
&& \alpha^{\prime} + A_r = 2\kappa e^{-U}\mathrm{Re}(e^{-i\alpha}\mathcal{L})\,, \nonumber \\
&& q^{\prime\,u} = \kappa e^{-U} h^{uv}\mathrm{Im}(e^{-i\alpha}\partial_v\mathcal L)\,, \nonumber \\
&& \ma Q^\prime = -4 e^{2\psi - 3U}\mathcal H\Omega\mathrm{Re}\tilde{\mathcal V}\,,
\label{eq:flowequations}
\end{eqnarray}
where also (\ref{eq:symcon}) and (\ref{eq:quantization}) must hold together with
\begin{equation}
2 e^U\mathcal H\Omega\mathrm{Re}\tilde{\mathcal V} = \ma H\Omega\ma A_t\,,
\label{eq:relalg}
\end{equation} 
since the last equ.~of \eqref{eq:flowequations} has to coincide with \eqref{eq:maxsim}. \eqref{eq:relalg}
is the symplectically covariant generalization of the constraint \eqref{eq:ReL-A}.

At the end of this subsection some comments on the limit of flat horizons ($\kappa=0$) are in order.
This case was not considered above, where we took $\kappa=\pm 1$ only. For $\kappa=0$, taking
(as in \cite{Cardoso:2015wcf}) $\ma P^1=\ma P^2=\ma Q^3=0$, one can again write the action
as a sum of squares, now with the Hamilton-Jacobi function $W=e^U|\mathcal Z-ie^{2(\psi-U)}\ma W^3|$.
The resulting first-order equations agree then, for purely electric gauging, precisely with those derived
in \cite{Cardoso:2015wcf}. (Note that the authors of \cite{Cardoso:2015wcf} considered electric gaugings
only, and did not identify the `superpotential' that drives their first-order flow).

\subsection{Non-BPS flow equations}

An interesting consequence of the flow equations in the Hamilton-Jacobi formalism is that the squaring of
the action is not unique; one can find another flow that squares the effective action in a similar way. This
was done for the ungauged case in \cite{Ceresole:2007wx} and for gauged supergravity with FI terms
in \cite{Gnecchi:2012kb}. We shall now generalize this procedure to the presence of hypermultiplets.

By repeating essentially the same computations as in the preceding subsection, one can show
that there is an alternative set of first-order equations that comes from the Hamilton-Jacobi function
\begin{equation}
W = e^U\left|\langle\tilde{\mathcal Q},\mathcal V\rangle + i\kappa e^{2(\psi - U)} \langle\ma W^x
\tilde{\ma Q^x},\mathcal V\rangle\right|\,, \label{eq:prW}
\end{equation}
with the associated constraints
\begin{equation}
\ma H\Omega\ma Q = 0\,, \qquad 2 e^U\mathcal H\Omega\mathrm{Re}\tilde{\mathcal V} = S\ma H
\Omega\ma A_t\,,
\label{eq:constraints}
\end{equation}
where we introduced a `field rotation matrix' $S\in\text{Sp}(2n_v+2,\mathbb R)$ that rotates the
charges as $\tilde{\ma Q} = S\ma Q$ and that has to satisfy the compatibility conditions
\begin{equation}
S\mathcal H S^T = \mathcal H\,, \qquad S^T\mathcal M S = \mathcal M\,. \label{eq:comp-cond}
\end{equation}
Moreover, the rotated charges must obey the analogue of \eqref{eq:quantization}, namely
\begin{equation}
\tilde{\ma Q^x}\tilde{\ma Q^x} =1\,.
\end{equation}
The first equ.~of \eqref{eq:constraints} is a consequence of spherical/hyperbolic symmetry, and implies,
together with $S\mathcal H S^T = \mathcal H$ and the fact that $S$ is symplectic,
the additional condition $\ma H\Omega\tilde{\ma Q}=0$. The latter and the equation
$\ma H\Omega\ma Q = 0$ lead respectively to
\begin{equation}
\langle\ma K^u,\tilde{\ma Q}\rangle = \langle\ma K^u,\ma Q\rangle = 0\,,
\label{eq:algebraic}
\end{equation} 
which are quite restrictive constraints on the possible gaugings. Moreover, in general it is not guaranteed
that a nontrivial solution to \eqref{eq:comp-cond} exists. Note that the technique of `rotating charges'
was first introduced in \cite{Ceresole:2007wx,LopesCardoso:2007qid}, and generalizes the sign-flipping
procedure of \cite{Ortin:1996bz}. It was applied to $\text{U}(1)$ FI-gauged supergravity
in \cite{Klemm:2012vm,Gnecchi:2012kb}.

\section{Attractors}
\label{sec:attractors}

The attractor mechanism \cite{Ferrara:1995ih,Strominger:1996kf,Ferrara:1996dd,
Ferrara:1996um,Ferrara:1997tw} has been the subject
of extensive research in the asymptotically flat case, and was extended more recently
in \cite{Morales:2006gm,Huebscher:2007hj,Bellucci:2008cb,Cacciatori:2009iz,Dall'Agata:2010gj,
Kachru:2011ps} to black holes with more general asymptotics. In particular, the authors
of \cite{Bellucci:2008cb} studied the attractor mechanism for $N=2$, $D=4$ gauged supergravity
in presence of $\text{U}(1)$ Fayet-Iliopoulos terms, and their results were extended
in \cite{Chimento:2015rra} to the case of hypermultiplets with abelian gaugings.
The attractor mechanism for a black hole solution describes the stabilization of the scalars on the event
horizon as a dynamical process of extremization of a suitable effective potential. This process is
completely independent of the initial values of the scalars, that flow to their horizon values which are
fixed by the black hole charges. The mechanism can be understood by studying the flow equations in the
near-horizon limit.

Following \cite{Dall'Agata:2010gj,Halmagyi:2013qoa,Erbin:2014hsa}, in this section we show that, 
in the near-horizon limit, the flow equations \eqref{eq:flowequations} become a set of algebraic equations
that determine the values of the vector scalars $z^i$ and the hyperscalars $q^u$ on the horizon in terms
of the charges and the gaugings and for this reason they are called attractor equations. 
As one can deduce from the general form of \eqref{eq:flowequations}, the results will be similar to those
obtained in \cite{Dall'Agata:2010gj,Halmagyi:2013qoa}, once we substitute the FI
parameters $\ma G$ by the expression $-\kappa\ma Q^x\ma P^x$.

\subsection{Attractor equations and near-horizon limit}

In order to derive the attractor equations, one has to make some assumptions on the behaviour of the
fields in the near-horizon limit, where we require all the fields and their derivatives to be regular.
To get the near-horizon geometry $\mathrm{AdS}_2\times\Sigma$ with
$\Sigma=\{\mathrm{S}^2,\mathrm{H}^2\}$, the warp factors must have the form
\begin{equation}
U = \log\left(\frac r{r_A}\right)\,, \qquad \psi = \log\left(\frac{r_S}{r_A}r\right)\,,
\end{equation}
where $r_A$ and $r_S$ denote the curvature radii of $\text{AdS}_2$ and $\Sigma$ respectively.
It is easy to show that $W=0$ at the horizon $r=0$; in fact the flow equations for $U$ and $\psi$ can be
rewritten as
\begin{equation}
U^\prime = -e^{-2(A+U)}(W - \partial_A W)\,, \qquad A^\prime = e^{-2(A+U)}W\,,
\end{equation}
where $A=\psi-U$ and $A\to\log(r_S)$ for $r\to 0$.
$W=0$ implies
\begin{equation}
\ma Z = -i\kappa r_S^2\ma L\,.
\end{equation}
Assuming $z^{\prime\,i}=0$ and $q^{\prime\,u}=0$ at the horizon, it follows that
\begin{equation}
D_i\ma Z = -i\kappa r_S^2 D_i\ma L\,, \qquad \mathrm D_u\ma L = 0\,,
\end{equation}
and $\alpha^\prime=0$. From $\mathrm{D}_u\ma L=0$ we get
\begin{equation}
\langle\ma K^v,\ma V\rangle = 0\,,
\label{eq:eqhyprNH}
\end{equation}
if we use also the algebraic relation $\langle\ma K^v,\ma Q\rangle=0$ (cf.~\eqref{eq:algebraic})
together with \eqref{eq:quaternionic_kahler_complexstruct_definition}, \eqref{eq:quaternionkahl_su2curv}
and \eqref{eq:mommaps}.
As in \cite{Chimento:2015rra}, we can choose the gauge $\ma A_t=0$ at the horizon. Then, from
\eqref{eq:relalg} and the last equation of \eqref{eq:flowequations}, one obtains $\ma Q^\prime=0$.

With these assumptions, the BPS flow equations \eqref{eq:flowequations} become
\begin{equation}
\begin{split}
&4\mathrm{Im}(\bar{\ma Z}\ma V) - \kappa r_S^2\Omega\ma M\ma Q^x\ma P^x + \ma Q = 0\,, \\
&\ma Z = -\frac{r_S^2}{2 r_A} e^{i\alpha}\,, \\
&\langle\ma K^v,\ma V\rangle = 0\,,
\label{eq:attractor}
\end{split}
\end{equation}
that must be supplemented by the constraints $\ma Q^x\ma Q^x=1$ and $\ma H\Omega\ma Q=0$.
If one rotates to a frame with purely electric gauging, $\ma Q^x$ boils down to
$p^\Lambda P^x_\Lambda$, and the magnetic charges $p^\Lambda$ become constant. One can then
use a local (on the quaternionic K\"ahler manifold) $\text{SU}(2)$ transformation to set
$\ma Q^1=\ma Q^2=0$, and the equations \eqref{eq:attractor} reduce to the ones obtained
in \cite{Erbin:2014hsa}.

The solutions of \eqref{eq:attractor} are the horizon values of the scalars in terms of the charges and the 
gaugings. Furthermore, taking in consideration homogeneous models and solving the attractor equations
for $r_S^2$, one can derive the Bekenstein-Hawking entropy written in \cite{Erbin:2014hsa} with the substitution $\ma P^3\to -\kappa\ma Q^x\ma P^x$. The main difference w.r.t.~the FI case consists in
the dependence of $\ma Q^x\ma P^x$ on the hypers, whose horizon values are fixed by
\eqref{eq:eqhyprNH} and by $\ma H\Omega\ma Q=0$.

\section{Examples of solutions}
\label{sec:examples}

The only known analytic black hole solution to $N=2$, $D=4$ gauged supergravity with running
hyperscalars was constructed in \cite{Chimento:2015rra}. In this section, we verify that this solves the flow
equations \eqref{eq:flowequations} and we consider a particular symplectic rotation of
the solution. Furthermore, we study some different gaugings of the universal hypermultiplet (UHM),
and obtain a family of black holes very similar to that of \cite{Chimento:2015rra}.

\subsection{\label{test}Test for the BPS flow}

The model considered in \cite{Chimento:2015rra} is defined by the prepotential $F=-i L^0 L^1$ and by
the universal hypermultiplet, i.e., the hyperscalars parametrize the quaternionic manifold
$\mathrm{SU}(2,1)/\mathrm{U}(2)$.

Using the hypermultiplet data given in \cite{Ceresole:2001wi}, the metric on the quaternionic manifold
reads\footnote{In our conventions the metric is rescaled by a factor of $1/2$ and the moment maps by a
factor of $2$ w.r.t.~\cite{Ceresole:2001wi}.}
\begin{equation}
h_{uv}\D q^u\D q^v = \frac{\D V^2}{4 V^2} + \frac1{4 V^2}(\D\sigma + 2\theta\D\tau - 2\tau\D\theta)^2
+ \frac1V(\D\theta^2 + \D\tau^2)\,.
\end{equation}
The gauging choosen in \cite{Chimento:2015rra} is defined by the Killing vectors $\vec k_1$ and
$\vec k_4$ of \cite{Ceresole:2001wi} such that
\begin{equation}
\mathcal P^x = \left(\begin{array}{c} 0 \\ c\delta^0_\Lambda P_4^x - k_\Lambda P_1^x \end{array}\right)\,.
\label{eq:gaugingsamu}
\end{equation}
Here $c$ and $k_\Lambda$ ($\Lambda=0,1$) denote constants, and $P_1^x$, $P_4^x$ are the
moment maps corresponding to $\vec k_1$, $\vec k_4$ respectively, that can be found
in \cite{Ceresole:2001wi}.

The Hamilton-Jacobi function driving the flow is given by
\begin{equation}
W = e^U\left|\langle\mathcal Q,\mathcal V\rangle + i \kappa e^{2(\psi - U)}\langle\ma W^x\ma Q^x,
\mathcal V\rangle\right|\,, \label{eq:prW}
\end{equation}
and the equations \eqref{eq:flowequations} must be solved together with the constraints
\eqref{eq:symcon} and \eqref{eq:quantization}. The latter immediately imply that the truncation
$\sigma=\tau =\theta=0$ is consistent. With this choice, and for $\kappa=-1$ (hyperbolic horizon), the
remaining nontrival components of \eqref{eq:symcon} and \eqref{eq:quantization} boil down to
\begin{equation}
p^0 k_0 + p^1 k_1 = 0\,, \qquad p^0 = \frac1c\,. \label{constr-charges-samuele}
\end{equation}
In presence of only magnetic charges, \eqref{eq:prW} becomes
\begin{equation}
W = e^U\left|\frac i{\sqrt{4z}}\left[p^0 z + p^1 - e^{2(\psi - U)}\left(c + \frac{k_0}{2V} + \frac{k_1z}{2V}
\right)\right]\right|\,,
\label{eq:Wsamu}
\end{equation}
where $z$ is the scalar field sitting in the vector multiplet.
Plugging \eqref{eq:Wsamu} into the BPS flow equations following from \eqref{eq:magsquaring} and
using appropriate ans\"atze for $U,\psi,z$ and the dilaton $V$, one recovers
\begin{equation}
\D s^2 = \frac{-4 p^1}{k_0} r^2\left[-\left(1 + \frac{k_0}{c r^2}\right)^2 r^2\D t^2 + \left(1 +
\frac{k_0}{c r^2}\right)^{-2}\frac{\D r^2}{r^2} + \frac12\D\Omega^2_{-1}\right]\,, \label{eq:solution-metr}
\end{equation}
\begin{equation}
z = \frac c{k_1} r^2\,, \qquad V = r^2\,, \qquad A^\Lambda =  p^\Lambda\sinh\theta\D\phi\,,
\label{eq:solution-scal-A}
\end{equation}
where the charges are constrained by \eqref{constr-charges-samuele}.
This is the black hole solution constructed in \cite{Chimento:2015rra}, where the parameters must satisfy
\begin{equation}
\frac {p^1}{k_0} < 0\,, \qquad \frac{k_0}c < 0\,, \qquad \frac c{k_1} > 0\,.
\end{equation}
These inequalities arise respectively from the requirements of having the correct signature, a genuine
horizon (at $r^2=-k_0/c$), and no ghosts in the action.

\subsection{Symplectic rotation of the electromagnetic frame}

One of the advantages of the symplectic covariance of the equations \eqref{eq:flowequations} is the
possibility of mapping solutions to solutions in different symplectic frames in presence of hypermultiplets,
as in the FI case \cite{Dall'Agata:2010gj}. Actually this was to be expected, since the hypermultiplets are
insensitive to electromagnetic duality rotations.

As an example, let us consider the mapping between the prepotentials $F=-iL^0L^1$ and
$F=\frac i4\tilde{L}^\Lambda\eta_{\Lambda\Sigma}\tilde{L}^\Sigma$, where
$\eta_{\Lambda\Sigma}=\text{diag}(-1, 1)$, and the reason for the different names for the upper parts of
the symplectic sections will become clear in a moment. The symplectic matrix \cite{Sabra:1996xg}
\[
T = \left(\begin{array}{cccc}
 1 & 1 & 0 & 0 \\
 1 & -1 & 0 & 0 \\
0 & 0 & \frac12 & \frac12 \\
0 & 0 & \frac12 & - \frac12
\end{array}\right)
\] 
realizes explicitly the isomorphism between the special K\"ahler structures described by these two prepotentials on the manifold $\text{SU}(1,1)/\text{U}(1)$. For the model with
$F=\frac i4\tilde{L}^\Lambda\eta_{\Lambda\Sigma}\tilde{L}^\Sigma$, the symplectic section reads 
\begin{equation}
\tilde{\mathcal V} = (\tilde{L}^0,\tilde{L}^1, -\frac i2\tilde{L}^0,\frac i2\tilde{L}^1)^T\,.
\end{equation}
Choosing the gaugings and the charge vector as
\begin{equation}
\tilde{\mathcal P}^x = \left(\begin{array}{c} 0 \\ \tilde{c}_\Lambda P_4^x - \tilde{k}_\Lambda P_1^x
\end{array}\right)\,, \qquad \tilde{\mathcal Q} = \left(\begin{array}{c} \tilde{p}^\Lambda \\ 0
\end{array}\right)\,,
\end{equation}
where $\tilde c_\Lambda$ and $\tilde k_\Lambda$ are constants, one can solve the BPS first-order flow driven by 
\begin{equation}
\tilde W = e^U\left|\langle\tilde{\mathcal Q},\tilde{\mathcal V}\rangle + i\kappa e^{2(\psi - U)}\langle
\tilde{\ma W}^x\tilde{\ma Q}^x,\tilde{\mathcal V}\rangle\right|\,,
\end{equation} 
using the solution \eqref{eq:solution-metr}, \eqref{eq:solution-scal-A} together with
\begin{equation}
\tilde{\mathcal V} = T\mathcal V\,, \qquad \tilde{\mathcal Q } = T\mathcal Q\,, \qquad
\tilde{\mathcal P}^x = T\mathcal P^x\,, \qquad \tilde{\mathcal G} = T\mathcal G\,.
\end{equation}
The solution in the rotated frame is given by the same metric and gauge fields of \eqref{eq:solution-metr},
\eqref{eq:solution-scal-A} (up to the redefinition of the parameters in $\mathcal Q $ and $\mathcal G$
in terms of the ones contained in $\tilde{\mathcal Q}$ and $\tilde{\mathcal G}$), but the vector
multiplet scalar is functionally modified to
\begin{equation}
\tilde z = \frac{1-z}{1+z}\,.
\end{equation}
As was to be expected, this is precisely the coordinate transformation from the metric of the Poincar\'e
disk,
\begin{equation}
\tilde{g}_{\tilde z\bar{\tilde z}} = \frac1{(1 - \tilde z\bar{\tilde z})^2}\,,
\end{equation}
to the one of the Poincar\'e upper half-plane,
\begin{equation}
g_{z\bar z} = \frac1{(z + \bar z)^2}\,.
\end{equation}

\subsection{Some different gaugings}

Consider the model of subsection \ref{test}, but with
a gauging defined by the Killing vectors $\vec k_4$ and $\vec k_6$ of \cite{Ceresole:2001wi}
such that
\begin{equation}
\mathcal P^x = \left(\begin{array}{c} 0 \\ c\delta^0_\Lambda P_4^x - k_\Lambda P_6^x\end{array}\right)\,,
\end{equation}
where $c$ and $k_\Lambda$ denote arbitrary constants and the moment maps $P_4^x$, $P_6^x$ are
given in \cite{Ceresole:2001wi}. Choosing the consistent truncation $\tau=\theta=\sigma=0$, one
obtains Hamilton's characteristic function
\begin{equation}
W = e^U\left|\frac i{\sqrt{4z}}\left[p^0 z + p^1 - e^{2(\psi - U)}\left(c + \frac{k_0}2 V + \frac{k_1}2 Vz
\right)\right]\right|\,,
\end{equation}
which is identical to \eqref{eq:Wsamu}, up to the substitution $V\rightarrow 1/V$, as also the truncated
moment maps show. Using an ansatz similar to the one in \cite{Chimento:2015rra}, it is easy to find a
new solution for this flow. $U$, $\psi$ and $z$ remain exactly the same as in \eqref{eq:solution-metr},
\eqref{eq:solution-scal-A}, but the dilaton becomes now
\begin{equation}
V = \frac1{r^2}\,.
\end{equation}

Another interesting isometry is $\vec k_5$ of \cite{Ceresole:2001wi}, i.e., the generator of dilatations.
Let us choose
\begin{equation}
\mathcal P^x = \left(\begin{array}{c} 0 \\ c\delta^0_\Lambda P_4^x - k_\Lambda P_5^x\end{array}\right)\,,
\end{equation}
together with the consistent truncation $\theta=\tau=0$, i.e., we keep two running hyperscalars
$V$ and $\sigma$. The Hamilton-Jacobi function is
\begin{equation}
W = e^U\left|\frac i{\sqrt{4z}}\left[p^0 z + p^1 - e^{2(\psi - U)}\left(c + \frac{k_0\sigma}{2V} +
\frac{k_1\sigma}{2V} z\right)\right]\right|\,.
\end{equation}
In this case the flow equations for the two hyperscalars can be brought to the form
\begin{equation}
\left(\begin{array}{c} V^\prime \\ \sigma^\prime \end{array}\right) =
-2\Omega\left(\begin{array}{c} V \\ \sigma \end{array}\right) k_\Lambda H^\Lambda\,, \qquad
H^\Lambda \equiv e^{-U} L^\Lambda\,, \label{eq:eqhypers}
\end{equation}
which imply
\begin{equation}
V^2 + \sigma^2 = \mbox{const}\,. \label{eq:hyperscalars}
\end{equation}
From this it is easy to see that the eqns.~\eqref{eq:eqhypers} decouple. In fact we get
\begin{equation}
V(r) = \rho(r)\cos\theta(r)\,, \qquad \sigma(r) = \rho(r)\sin\theta(r)\,,
\end{equation}
where
\begin{equation}
\rho^{\prime} = 0\,, \qquad \theta^\prime = -2 k_\Lambda H^\Lambda\,.
\end{equation}
The equation for $\theta$ is the same as the one for the hyperscalar with the
gauging \eqref{eq:gaugingsamu}, but unfortunately the eqns.~for $U$ and $z$ are different, and thus
\eqref{eq:solution-metr}, \eqref{eq:solution-scal-A} is not a solution for this gauging.

\section{Final remarks}
\label{sec:final}

In this paper, we considered $N=2$ supergravity in four dimensions, coupled to an arbitrary number
of vector- and hypermultiplets, where abelian isometries of the quaternionic hyperscalar target
manifold are gauged. For a static and spherically or hyperbolically symmetric ansatz, we derived
a system of first-order flow equations by making essential use of the Hamilton-Jacobi formalism.
We then included also magnetic gaugings and generalized our results to a symplectically covariant
form as well as to the non-BPS case. Moreover, as an immediate application of our first-order system,
we obtained the symplectically covariant attractor equations for gauged supergravity with both
vector- and hypermultiplets. Finally, some explicit black hole solutions with running hyperscalars
were given for a model containing the universal hypermultiplet plus one vector multiplet, for several
choices of gaugings. We hope that the results presented here will contribute to a more systematic
study of black holes in gauged supergravity with hypermultiplets; a topic on which little is known
up to now. Let us conclude our paper with the following suggestions for possible extensions and
questions for future work:
\begin{itemize}
\item Try to solve the flow equations \eqref{eq:flowequations} for models more complicated than the
one in \cite{Chimento:2015rra}.
\item Extend them to the nonextremal case by modifying Hamilton's characteristic function,
similar in spirit to what was done
in \cite{Miller:2006ay,Cardoso:2008gm,Andrianopoli:2009je,Gnecchi:2014cqa}.
\item Extend them to the rotating case and to other dimensions.
\item In the case where the scalar manifolds have some special geometric properties (e.g.~symmetric),
it may be possible to classify the attractor points as was done for ungauged supergravity
in e.g.~\cite{Bellucci:2006xz}.
\end{itemize}
We hope to come back to these points in a forthcoming publication.

\section*{Acknowledgements}

This work was supported partly by INFN. We would like to thank A.~Marrani and I.~Papadimitriou
for useful discussions.

\end{document}